\documentstyle[12pt,epsfig]{article}
\begin{document}

\begin{center}
{\bf RESONANT STATES IN $~^3 P$ CHANNELS OF CHARMONIUM}

E. Di Salvo, M. Pallavicini and E. Robutti \\

{\small \it
Dipartimento di Fisica dell'Universit\`a di Genova - INFN, Sezione di 
Genova, \\ Via Dodecaneso 33, 16146 Genova, Italy.}
\end {center}
\vspace {1.0 cm}      
\normalsize
\begin{abstract}
We employ QCD sum rules, implemented with two numerical algorithms already 
tested in two different channels of Charmonium, in order to predict masses of
resonances just above the ground states in the $~^3 P$ channels. We find that 
such masses are above the threshold of open charm. We calculate also the 
partial decay widths of the ground states into light hadrons and, for even 
spins, into two gammas; we find consistency with data.
\end{abstract}
\centerline{PACS 11.50.Li, 14.40.Jz}

\centerline{To appear in Physics Letters B} 
\vskip .30in
\baselineskip=1.2\baselineskip
A lot of experimental and theoretical efforts have been 
performed in order to study hadrons formed by one or more heavy
quarks. In particular, as regards spectroscopy predictions, essentially 
three different approaches have been adopted in the literature: $i)$ 
potential models\cite{rp,rpp,li}; $ii)$ lattice calculations\cite{ek}; 
$iii)$ QCD sum rules\cite{r1,r2,r3,r4}. Concerning 
heavy quarkonia, a systematic study of spectra has been done on 
Bottomonium\cite{r4p,r4s}, but not on Charmonium. We think the gap can be 
partially filled in by means of power moment QCD sum 
rules\cite{r1,r2,r4,r5,r6}, which, incidentally, in the case of Charmonium 
appear more suitable than exponential sum rules\cite{r3,r4p,r4s,r6t}. These 
have been already used successfully, with the help of 
two rather complex numerical algorithms, in the $~^1 S_0$ and $~^1 P_1$ 
channels\cite{r5,r6} of Charmonium, in order to determine the mass of the 
first excited state - be it a bound state ($\eta'_c$, $~^1 S_0$ channel)
or a resonant state ($h'_c$, $~^1 P_1$ channel) - and, as a
byproduct, some partial decay widths of ground states. In this letter we
extend our investigation to the $~^3 P$-channels of Charmonium; as a result we 
find that they have no bound states besides the ground one. Moreover we 
calculate the partial decay widths of the ground states into light hadrons and
(for even spin resonances) into gammas, finding consistency with data.

We recall shortly the power moment QCD sum rules, together with the two
numerical algorithms, which we have elaborated in preceding papers\cite{r5,r6}.
We assume a dispersion relation for the Fourier transform of the
two-point function of the current (or density) of the charmed quarks with the
quantum numbers of the channel considered. We consider spacelike overall
momenta, $q^2$, therefore we may develop the 
two-point function according to a Wilson expansion - a nontrivial 
one, including the contribution of the lowest-dimensional gluon condensate. 
Then for a given channel $\Gamma$ and for any positive integer $n$ we 
get\cite{r5}  

\begin{equation} 
A^{\Gamma}_n [1+\alpha_s a_n^{\Gamma}(\xi) +\Phi b_n^{\Gamma} (\xi)] =   
\ {1 \over \pi} \int \limits_0^{\infty} {{Im \Pi^{\Gamma} (s)} \over 
{(s+Q^2)^{n+1}}} ds,
\label{rul}
\end{equation}

where $Q^2$ = $-q^2$, $\alpha_s$ = $\alpha_s[4(m_c^0)^2+Q^2]$ is the running 
coupling constant ($\alpha_s[4(m_c^0)^2] = {\overline \alpha_s} \sim 0.3$), 
$\xi = {{Q^2} \over {4(m^0_c)^2}}$, $m^0_c = m_c(p^2=-m_c^2)$, 

\begin{equation}
\Phi = {{\pi^2} \over {36(m_c^0)^4}} 
<0|{{\alpha_s} \over {\pi}} G^a_{\mu\nu}G_a^{\mu\nu}|0>
\end{equation}

and $G^a_{\mu\nu}$ is the QCD strength tensor field. The l.h.s. of eq.
(\ref{rul}) coincides, up to an $n$-dependent factor, with the $n$-th moment of
the Wilson expansion. Concerning the spectral function, which appears at the 
r.h.s. of eq. (\ref{rul}), for heavy quarkonia it may be parametrized as 
follows:

\begin{equation}
Im \Pi^{\Gamma} (s) ={{9 \pi} \over 4}  \sum_{i=1}^N {{m_i^2} \over {g_i^2}} 
\delta (m_i^2-s) + {\sigma_0^{\Gamma} \over {8 \pi}} (1+ 
{{\overline \alpha_s} \over {\pi}}) \theta(s-s_0), 
\label{spec}
\end{equation}

where $m_i$ is the mass of the $i$-th resonance of the channel 
and $g_i$ the coupling constant of that resonance to the current (or density)
relative to the channel, while $s_0$ is 
the threshold energy squared of continuum; lastly 
$\sigma_0^{\Gamma}$ is a positive integer, which we set equal
to 1 for $P$-channels\cite{r6}. Owing to duality, we are free to assume 
$N$ = 1 or 2 in the sum that appears at the l.h.s. of eq. 
(\ref{spec}), including the other resonances in the continuum contribution. 
If we take $N$ = 1, from eqs. (\ref{rul}) and (\ref{spec}) we deduce the mass
of the ground state, i. e.,

\begin{equation}
m_1^2 = {{\overline M^{\Gamma}_n} \over {\overline M^{\Gamma}_{n+1}}}  - 
\ Q^2, \ ~~~~~ \ {\overline {M_n}^{\Gamma}} = {M_n}^{\Gamma} -  
{{\sigma_0^{\Gamma} (1+ {{\overline \alpha_s} \over {\pi}})} \over 
\ {8\pi^2 n (s_0+Q^2)^n}},
\label{emone}
\end{equation}

where $M_n^{\Gamma} = A^{\Gamma}_n [1+\alpha_s a_n^{\Gamma}(\xi) +\Phi 
\ b_n^{\Gamma} (\xi)] $ is the l.h.s. of eq. (\ref{rul}). If, instead, we take
$N$ = 2, we get the mass of the first excited state, i. e., 

\begin{equation}
m_2^2 = {{N^{\Gamma}_n} \over {N^{\Gamma}_{n+1}}} - Q^2, \ ~~~~~ \ 
N^\Gamma_n = \overline M^\Gamma_n - {{9 m_1^2} \over {4 g_1^2 
(m_1^2 + Q^2})^{n+1}} .  
\label{emtwo}
\end{equation}

Equations (\ref{emone}) and (\ref{emtwo}) yield respectively $m_1$ and $m_2$
as functions of $n$, $Q^2$, ${\overline \alpha_s}$, $\Phi$, $m^0_c$ and $s_0$, 
the last parameter varying from channel to channel. $m_{2}$ depends also on 
$g_1$, which is not known for most channels. Preceding analyses yield $\Phi$ = 
$1.35 ~ 10^{-3}$\cite{r2,r3}, ${\overline \alpha_s}$ = $0.3$\cite{r3,r4} and 
$m_c^0$ = $1.268 ~ GeV$\cite{r5}. Moreover, as we have already shown\cite{r5},
we can fix criteria for determining the other parameters. 

In principle the value of $s_0$ that appears in formula (\ref{emone}) could be 
different from the one used in formula (\ref{emtwo}), since in the latter case 
we exclude the contribution of $m_2$ from the continuum of 
the spectral function. However eq. (\ref{spec}) is a simplified 
parametrization of the real spectral function, which presents narrow resonances 
at low energies and therefore has a greater weight at low $s$ in the integral
at the r.h.s. of eq. (\ref{rul}). Then $s_0$ results to be a decreasing
function of $n$ and $Q^2$. We can set $s_0^{(1)}$ = 
$s_0^{(2)}$, provided $n_1 \geq n_2$ and $Q^2_{(1)} > Q^2_{(2)}$, 
the index 1 (2) referring to the option $N = 1 ~ (2)$. This self-consistency 
condition is always respected, both in the present analysis (see below) and in
preceding ones\cite{r5,r6,r6s,r6v}.   

In order to fix $n$ and $Q^2$, first of all we consider the dependence of
$m_{1(2)}$ on $n$ at a fixed $Q^2$. $m_i(n)$ present minima at given 
${\overline n}_i$, $i = 1, 2$, and therefore plateaux around such values (see 
e. g. fig. 1). Furthermore the amplitude of each plateau, 
which can be measured by the inverse of the quantity  

\begin{equation}
D_i = m_i({\overline n}_i - k_i) + m_i({\overline n}_i + k_i) - 
\ 2m_i({\overline n}_i),
\label{der}
\end{equation}

($k_1$ = 2, $k_2$ = 1), varies with $Q^2$. We choose $\xi_i$ so
as to coincide with locations of minima of $D_i$ (see figs. 2). In particular 
we pick up for $\xi_1$ the lowest local minimum (at least for not too large 
$\xi$) of $D_1$, provided the corresponding $m_1$ is stable with respect to the 
parameters $m_c^0$ and $s_0$; as we shall see below, the stability condition 
in the $~^3 P_1$ channel requires a particular care. As to 
$\xi_2$, we pick up the lowest local minimum of $D_2$ which does not coincide 
with any minima of $D_1$.

We determine $s_0$ by equalling $m_1$ to the mass of the ground state, which is
generally known experimentally. In order to determine $g_1$, we proceed as in
the vector channel\cite{r5}, considering the graph of 
${\overline D}_2$ (that is, of the value of $D_2$ obtained by selecting the
minimum in the way just described) versus $g_1$. The value corresponding to an 
oblique inflexion (fig. 3) is assumed  to be the 
correct one, provided it is stable with respect to small variations of the
parameters. This in turn allows to determine $m_2$ by the formulae exposed 
above.

\begin{table*}[hbt]
\caption{Results of the analysis in different channels of Charmonium}
\label{tab:param}
\begin{tabular*}{\textwidth}{@{}l@{\extracolsep{\fill}}ccccc}
\hline
                 & \multicolumn{1}{c}{$m_2 (MeV)$} 
                 & \multicolumn{1}{c}{$~~~g_1~~~$} 
                 & \multicolumn{1}{c}{$s_0 (GeV)^2$} 
                 & \multicolumn{1}{c}{$~~~~n~~~~$} 
                 & \multicolumn{1}{c}{$Q^2 ~ (GeV)^2$}    \\     
\hline
$~^3 P_0$  & $ ~ \ ~ \ 4097^{+32}_{-44}$ & $13.38^{+0.34}_{-0.07}$ & $23.7^{+0.12}_{-0.16}$ & $9 (6)$ & $15.4 (8.4) $ \\
$~^3 P_1$  & $ ~ \ ~ \ 4327^{+40}_{-32}$ & $16.0^{+0.5}_{-0.7}$    & $19.7^{+1.2}_{-0.9}$   & $8 (6)$ & $18.7 (13.4)$ \\
$~^3 P_2$  & $ ~ \ ~ \ 4466^{+34}_{-21}$ & $11.82^{+0.43}_{-0.32}$ & $20.8^{+1.3}_{-0.8}$   & $8 (6)$ & $21.5 (15.8)$ \\
\hline
\end{tabular*}
\end{table*}

Table 1 resumes the results of our analysis in the $~^3 P$-channels.
In particular values of $n$ and $Q^2$ outside (inside) parentheses refer to 
the option $N$ = 1 (2) in eq. (\ref{spec}).
The values of $m_2$ are above the threshold of open charm (3727 $MeV$),
therefore according to our predictions they are resonant
states, which decay into charmed particles. We observe that the above mentioned
self-consistency condition is satisfied for all $~^3 P$-channels. Another 
condition, necessary in order to check stability of minima of 
$D_1$ with respect to parameters, consists in examining the two graphs 
corresponding to eqs. $m_1(m^0_c, s_0)$ = 
$m_{GS}$,  $m_2(m^0_c, s_0)$ = $m_{FR}$, where $GS$ means "ground state" and 
$FR$ "first resonant state". Such graphs, represented in figs. 4 and 5, are 
quite similar to those drawn for the vector channel\cite{r5}, to the 
difference that in the vector channel both $m_{GS}$ and $m_{FR}$ are known 
experimentally, whereas in the other channels $m_{FR}$ is deduced from our 
numerical procedure. The analogous behaviour witnesses the self-consistency of 
our method. This is why in the $~^3 P_1$-channel we have not chosen the lowest 
minimum of $D_1$. Indeed this minimum - which yields erratic values of some 
parameters, like $\xi_i$ and $s_0$, and a mass $m_2$ greater than the one of
the first resonance of the $~^3 P_2$-channel - corresponds to a pair of
graphs quite different from the others, as one can see in fig. 5b.

Our mass predictions differ from others deduced on
the basis of potential models\cite{r7}; on the contrary they fulfill, within 
errors, the center-of-gravity rule derived for $P$-channels of Charmonium 
(see\cite{li} and refs. therein). 

We have already shown\cite{r5,r6} that $g_1$ contains useful information for 
predicting some partial decay widths. The formulae for decay withs
of $\chi_{c0(2)}$ into two gammas (gluons) read

\begin{equation}
\Gamma_{\chi \rightarrow 2 \gamma} = \frac{4}{3}
\frac{2 \pi {\overline g}^2}{(2J+1) m} \left(\frac{2}{3}\right)^4 \alpha^2, 
\ ~~~~~ \ \Gamma_{\chi \rightarrow 2 g} = \frac{9}{8} 
\left(\frac{\alpha_s}{\alpha}\right)^2 \Gamma_{\chi \rightarrow 2 \gamma} 
\label{wid}
\end{equation}

where $m$ = $m_{GS}$, $J$ ($0$ or $2$) the spin of the resonance,

\begin{equation}
{\overline g}^2 = \frac{9 m^{2l}}{4 g_1^2 f T}, \ ~~~ \
T = \frac{1}{4 m_c^2} Tr [(\rlap/k - \rlap/q + m_c) {\tilde O} 
(\rlap/k + m_c) O ],
\label{tr}
\end{equation}

$k$ is the four-momentum of the quark, $l$ an integer, $O$ an operator and $f$
an $O$-dependent normalization factor. For $~^3 P_{0(2)}$
channels $T$ turns out to depend critically on the modulus squared of the 
relative four-momentum $p$ $=$ $2k-q$ of the quark with respect to the 
antiquark. The average of $p^2$ over the hadronic state results to be $m^2 - 
4m_c^2$. According to Novikov et al.\cite{r1} (see also ref.\cite{r6}), we 
assume $p^2$ = $-m_c^2$, therefore $m_c^2$ = $\frac{1}{3} m^2$. 
In the $~^3 P_0$-channel $l_{\chi_{c0}}$ = 1, moreover $O_{\chi_{c0}}$ is the 
identity operator, yielding $f$ = 1 and $T_{\chi_{c0}} \simeq 0.268$.
As to the $~^3 P_2$-channel, $l_{\chi_{c2}}$ = 2 and

\begin{equation}
O_{\chi_{c2}} = \gamma_{\mu} p_{\nu} + \gamma_{\nu} p_{\mu} + \frac{2}{3} 
\ \eta_{\mu\nu} \rlap/p, \ ~~~ ~~~ \
\ \eta_{\mu\nu} = \frac{q_{\mu} q_{\nu}}{q^2} - g_{\mu\nu},
\label{op}
\end{equation}

Hence $T_{\chi_{c2}}$ = $\frac{46}{3}m_c^2$ and $f$ = 1/5. Calculations yield 

\begin{eqnarray}
\Gamma_{\chi_{c0} \rightarrow 2 \gamma} &=& 14.1^{+0.14}_{-0.7} 
\ ~ (5.6^{+6.4}_{-4.1}) ~ keV,
\\
\Gamma_{\chi_{c0} \rightarrow 2 g} &=& 26.8^{+0.3}_{-1.3} ~ (14.0\pm 5) ~ MeV, 
\\
\Gamma_{\chi_{c2} \rightarrow 2 \gamma} &=& 
\ 0.99\pm 0.09 ~ (0.32^{+0.14}_{-0.12}) ~ keV,
\\  
\Gamma_{\chi_{c2} \rightarrow 2 g} &=& 1.88^{+0.11}_{-0.13} ~ 
\ (1.73\pm 0.16) ~ MeV,
\end{eqnarray}

parentheses reporting experimental values\cite{rpa}.

As regards $\chi_{c1}$, we assume, according to Novikov et al.\cite{r1}, the 
decay into light hadrons to be dominated by the two-step
process $\chi_{c1}$ $\rightarrow$ $g ~ g^*$, $g^*$  $\rightarrow$
$q {\overline q}$, i. e.,

\begin{equation}
\Gamma_{\chi_{c1} \rightarrow gq{\overline q}} = \frac{1}{2m} 
\ {\overline g}^2_{\chi_{c1}} 
\ \frac{4}{3} (4\pi\alpha_s)^3 \int d\Phi_3 \frac{1}{\tilde{q}^4} 
\ \frac{1}{4} h_{\mu\nu} H^{\mu\nu},   
\label{wi1}
\end{equation}

where $d\Phi_3$ is the phase space of the $g q {\overline q}$ system,
$h_{\mu\nu}$ ($H_{\mu\nu}$) the hadronic tensor of light (charmed) quarks and
$\tilde{q}^2$ the effective mass squared of the $q {\overline q}$ system.
Furthermore ${\overline g}_{\chi_{c1}}^2$ is given by eqs. (\ref{tr}), with 
$l_{\chi_{c1}}$ = 2 and $O_{\chi_{c1}}$ = $\gamma_5 \gamma_{\mu}$, whence 
$T_{\chi_{c1}}$ = 1 and $f$ = 1. Then eq. (\ref{wi1}) yields

\begin{equation}
\Gamma_{\chi_{c1} \rightarrow gq{\overline q}} = 
\ \frac{4}{3} ~ \frac{12 m^6 \alpha_s^3}{g_1^2 m_c^2}
\ \int_{-1}^{1} d cos\theta \int_{\epsilon}^{0.5} dx ~ \omega ~ x^2 
\ \frac{1-cos\theta cos\phi}{\tilde{q}^4},
\end{equation}

where $\epsilon = \delta/[1+\delta-cos\theta(1-\delta)]$, $\delta = q_0^2/m^2$,
$q_0^2$ is a lower limit to $\tilde{q}^2$, $\omega$ the energy of the "real" 
gluon, $\theta$ 
($\phi$) the angle between the direction of the real gluon and the one of the
(anti-) quark in the overall cms. $\tilde{q}^2$, $\omega$ and $\phi$ depend on 
$\theta$  and $x$ through
four-momentum conservation constraints. Taking into account the value of $g_1$
(see table 1), and assuming, according to the lower bound of perturbative QCD
scales, $q_0^2$ = $4$ - $5$ $GeV^2$, we get $\Gamma_{\chi_{c1} \rightarrow 
gq{\overline q}}$ = ($658_{-207}^{+131}$) ~ $keV$, quite consistent with the 
experimental data, ($640\pm100$) ~ $keV$.

Other authors\cite{r8,r9,r10} calculate partial decay widths of Charmonium
$P$-wave resonances by means of the factorization theorem, including colour 
octet contributions in order to eliminate infrared divergences\cite{r11}. They
do not calculate the nonperturbative factors involved, rather they determine
them either from some of the partial decay widths or from low-energy $e^+ - 
e^-$ data\cite{r12}. On the contrary, we do not take into account the octet 
contribution (relatively small in percentage); however, thanks to QCD sum 
rules, we can calculate nonperturbative factors.

\vfill
\eject

\begin{figure}
\leavevmode
\centering
\epsfig{file=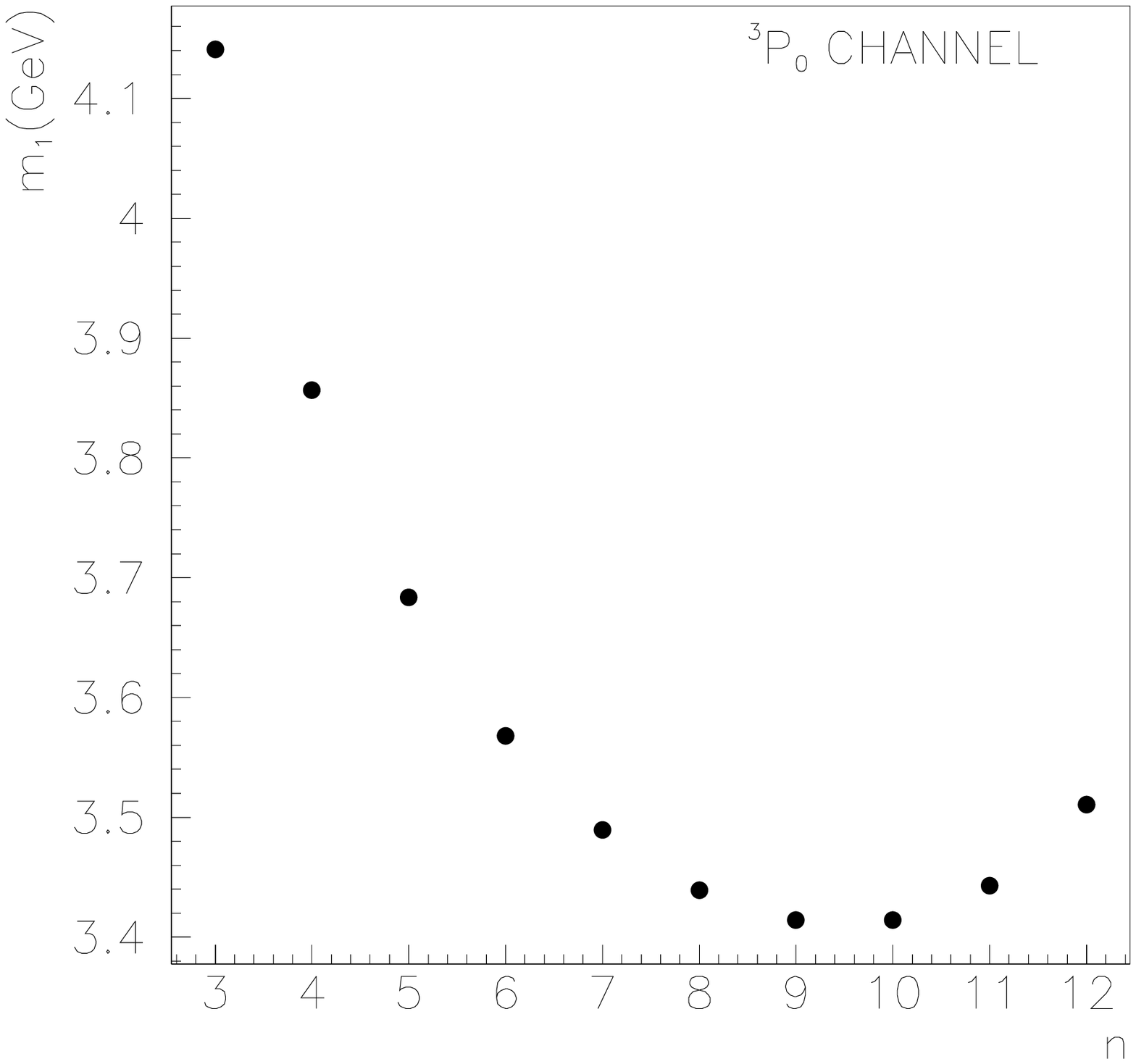,width=7cm}
\caption{$~^3 P_0$ channel: behaviour of $m_1$ vs $n$ for $\xi = 2.5$ }
\label{fig:one}
\end{figure}

\begin{figure}
\leavevmode
\centering
\parbox{7cm}{
\leavevmode
\centering
\epsfig{file=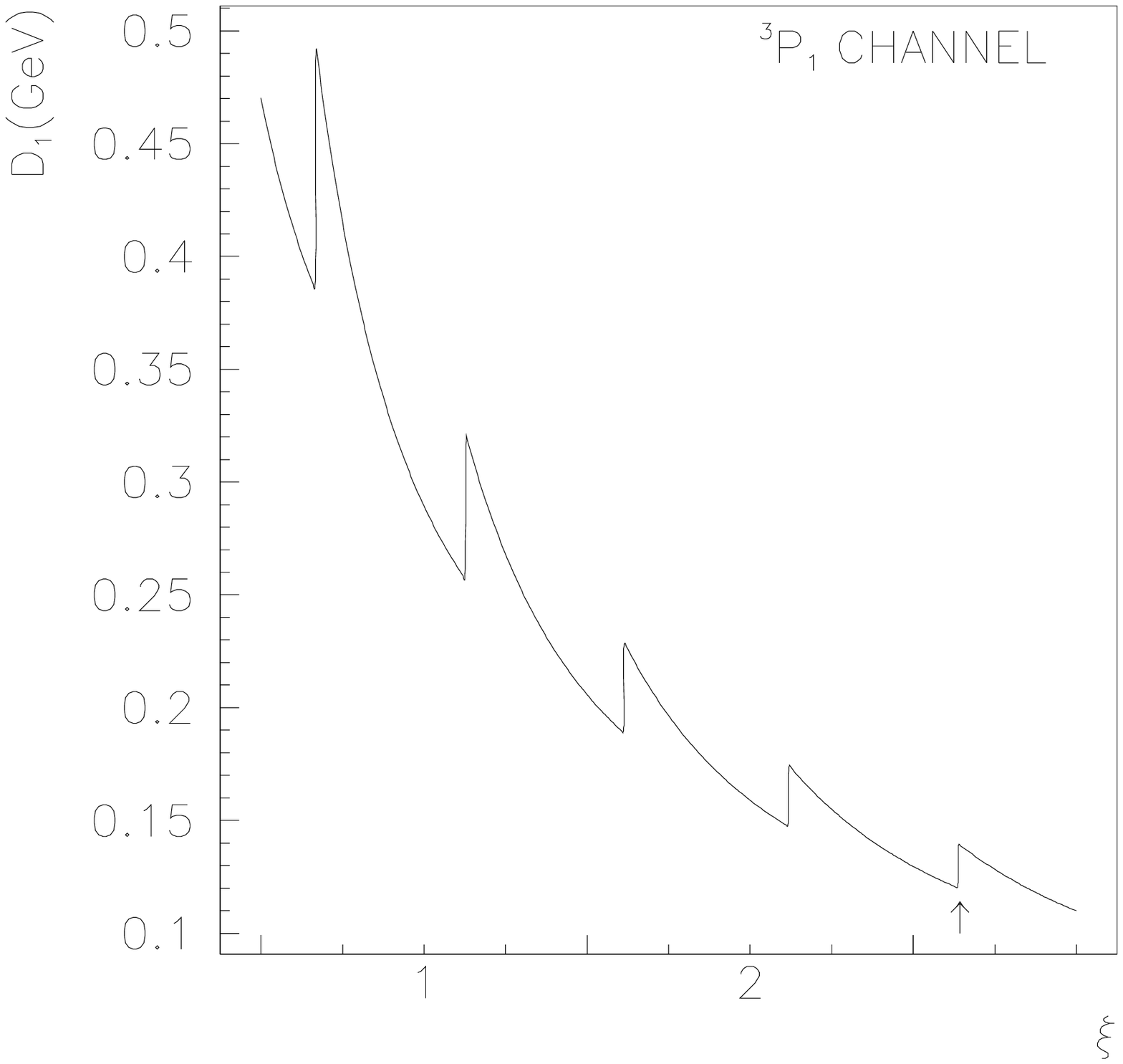,width=7cm}
\\
$a)$
}
\hspace{1cm}
\parbox{7cm}{
\leavevmode
\centering
\epsfig{file=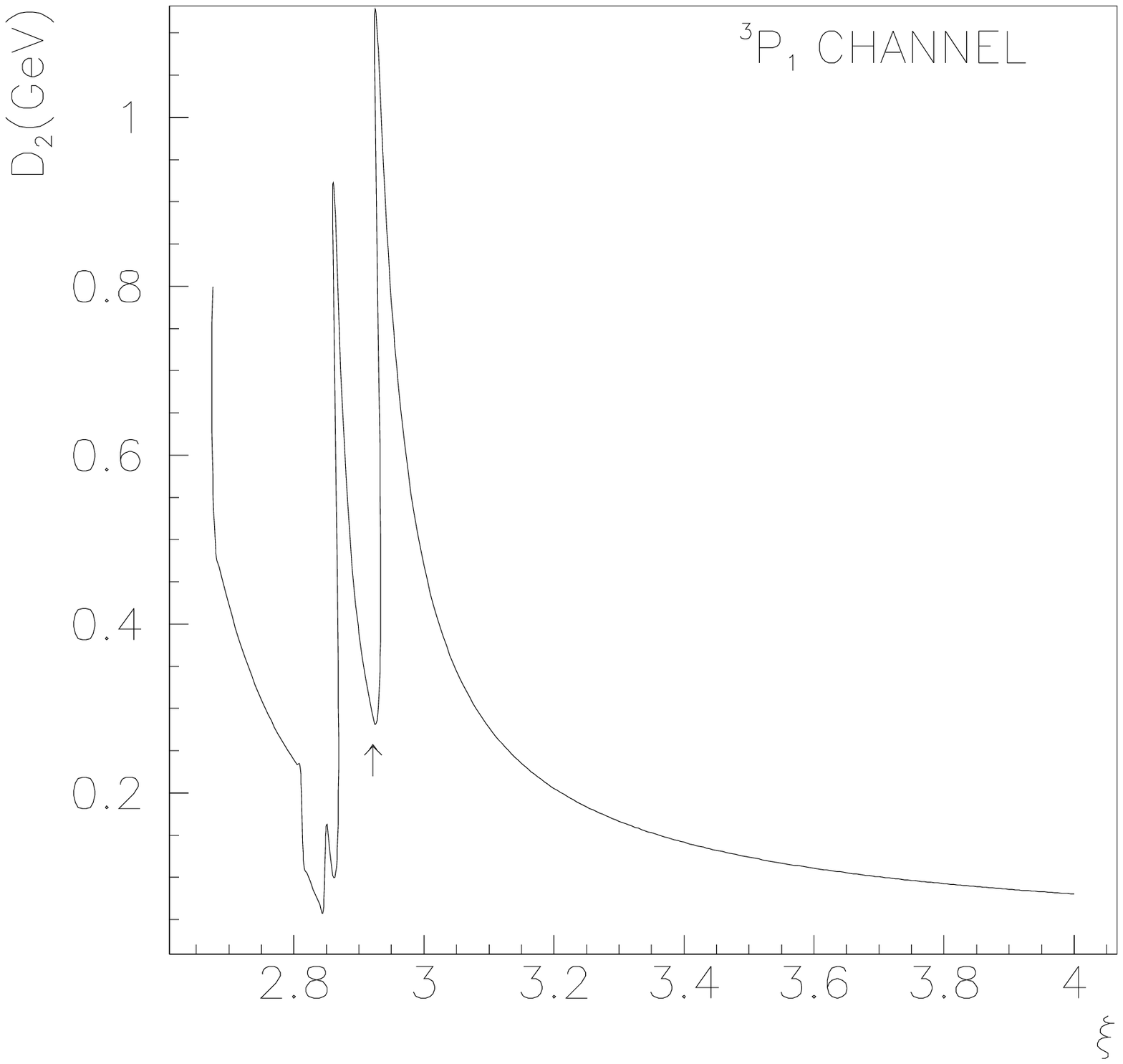,width=7cm}
\\
$b)$
}
\caption{$~^3 P_1$ channel: behaviour of $a)$ $D_1$ and $b)$ $D_2$ vs $\xi$} 
\label{fig:two}
\end{figure}

\begin{figure}
\leavevmode
\centering
\epsfig{file=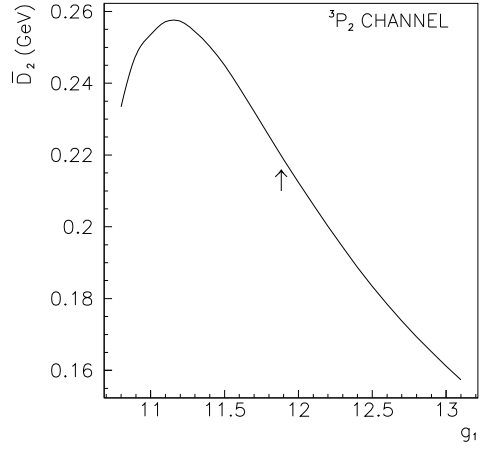,width=7cm}
\caption{$~^3 P_2$ channel: ${\overline D}_2$ vs $g_1$. The arrow indicates the 
oblique inflexion.}
\label{fig:three}
\end{figure}

\begin{figure} 
\leavevmode 
\centering 
\parbox{7cm}{ 
\leavevmode 
\centering
\epsfig{file=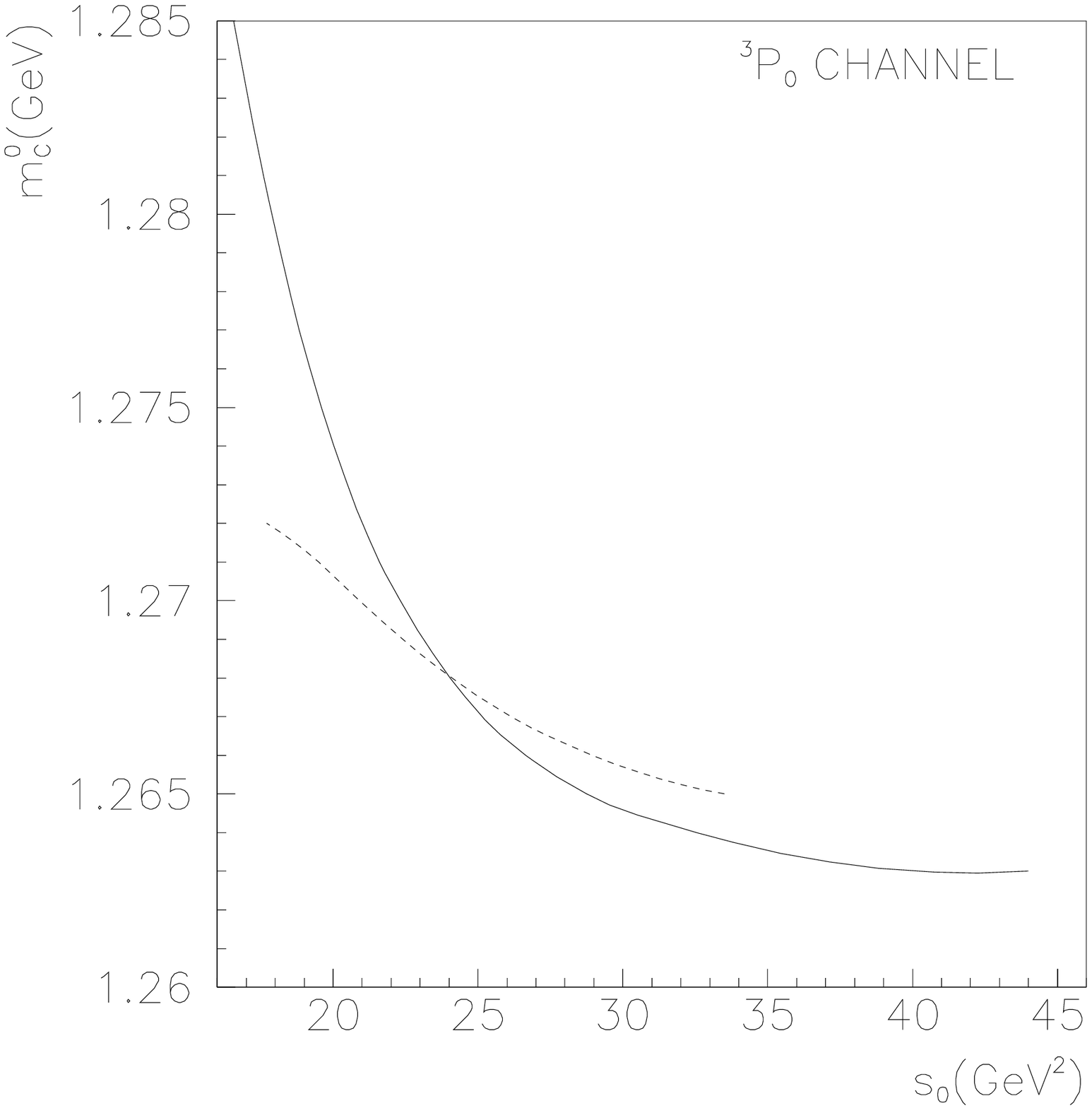,width=7cm} 
\\ 
$a)$ 
} 
\hspace{1cm} 
\parbox{7cm}{
\leavevmode 
\centering 
\epsfig{file=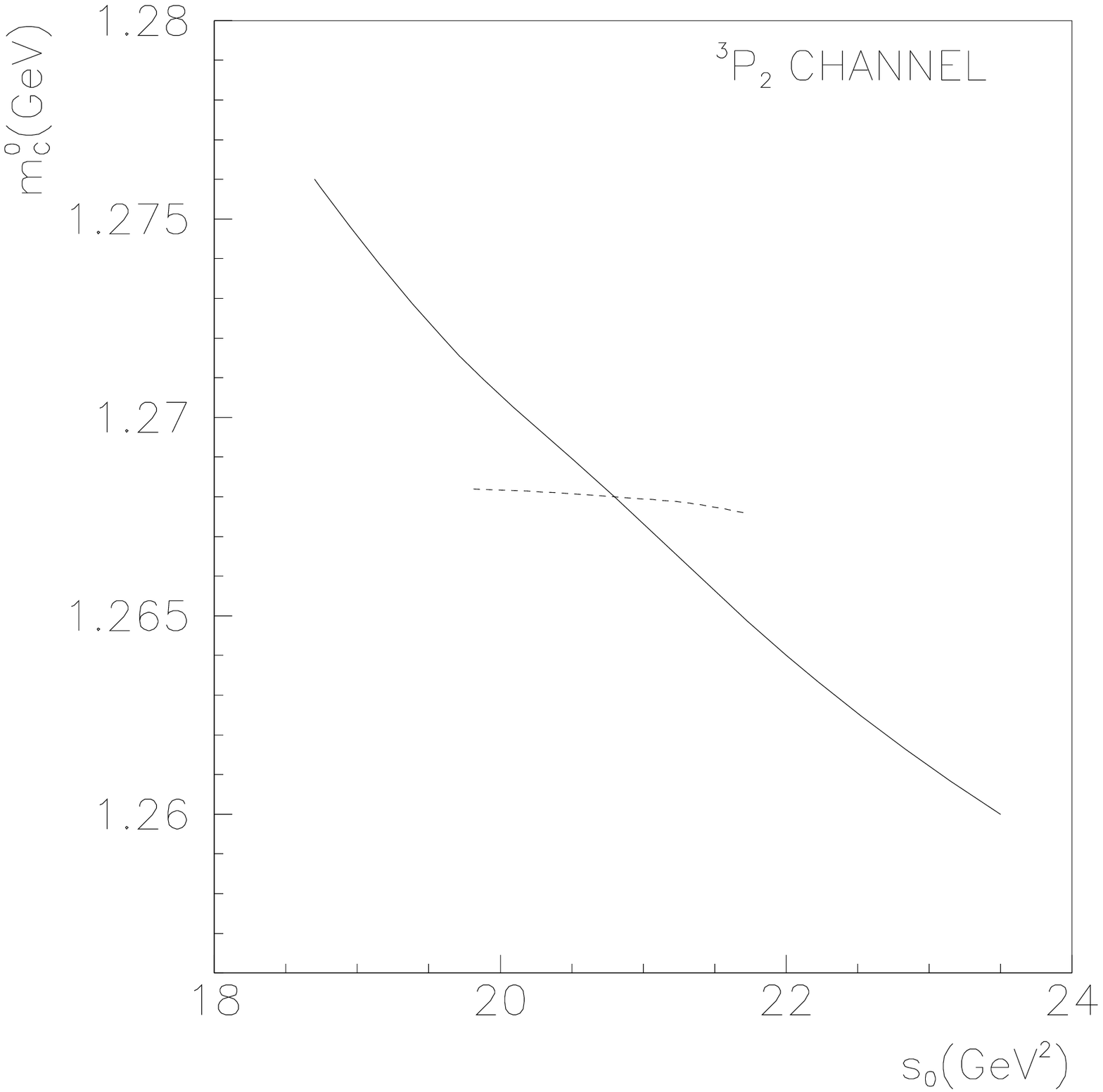,width=7cm} 
\\ 
$b)$ 
}
\caption{$~^3 P_0$ and $~^3 P_2$ channels: $m_c^0$ vs $s_0$. The full lines 
refer to eq. $m_1 = m_{GS}$, the dashed ones to eq. $m_2 = m_{FR}$, 
where $GS$($FR$) means "ground state" ("first resonant state").} 
\label{fig:four}
\end{figure}

\begin{figure} 
\leavevmode 
\centering 
\parbox{7cm}{ 
\leavevmode 
\centering
\epsfig{file=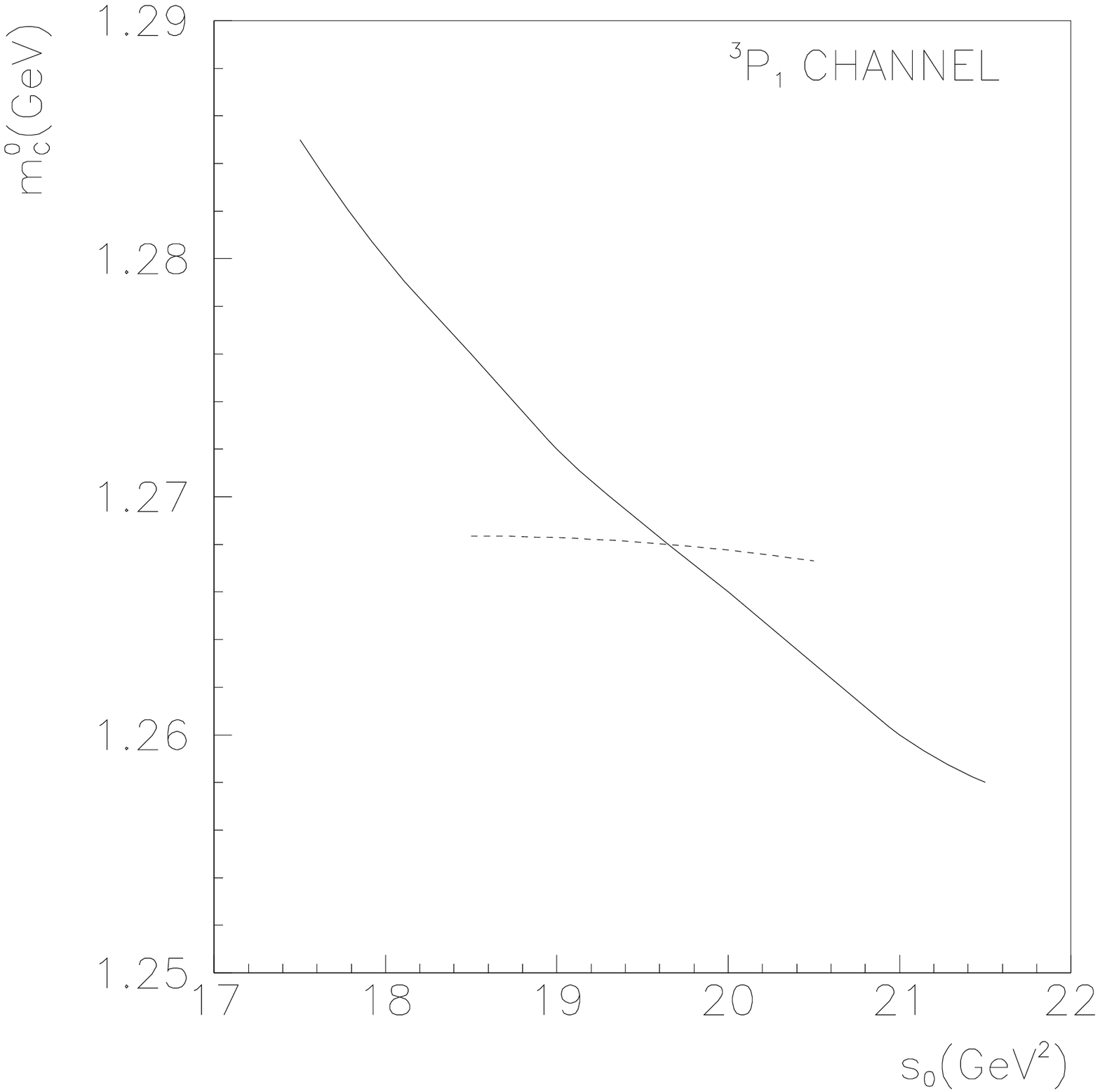,width=7cm} 
\\ 
$a)$ 
} 
\hspace{1cm} 
\parbox{7cm}{
\leavevmode 
\centering 
\epsfig{file=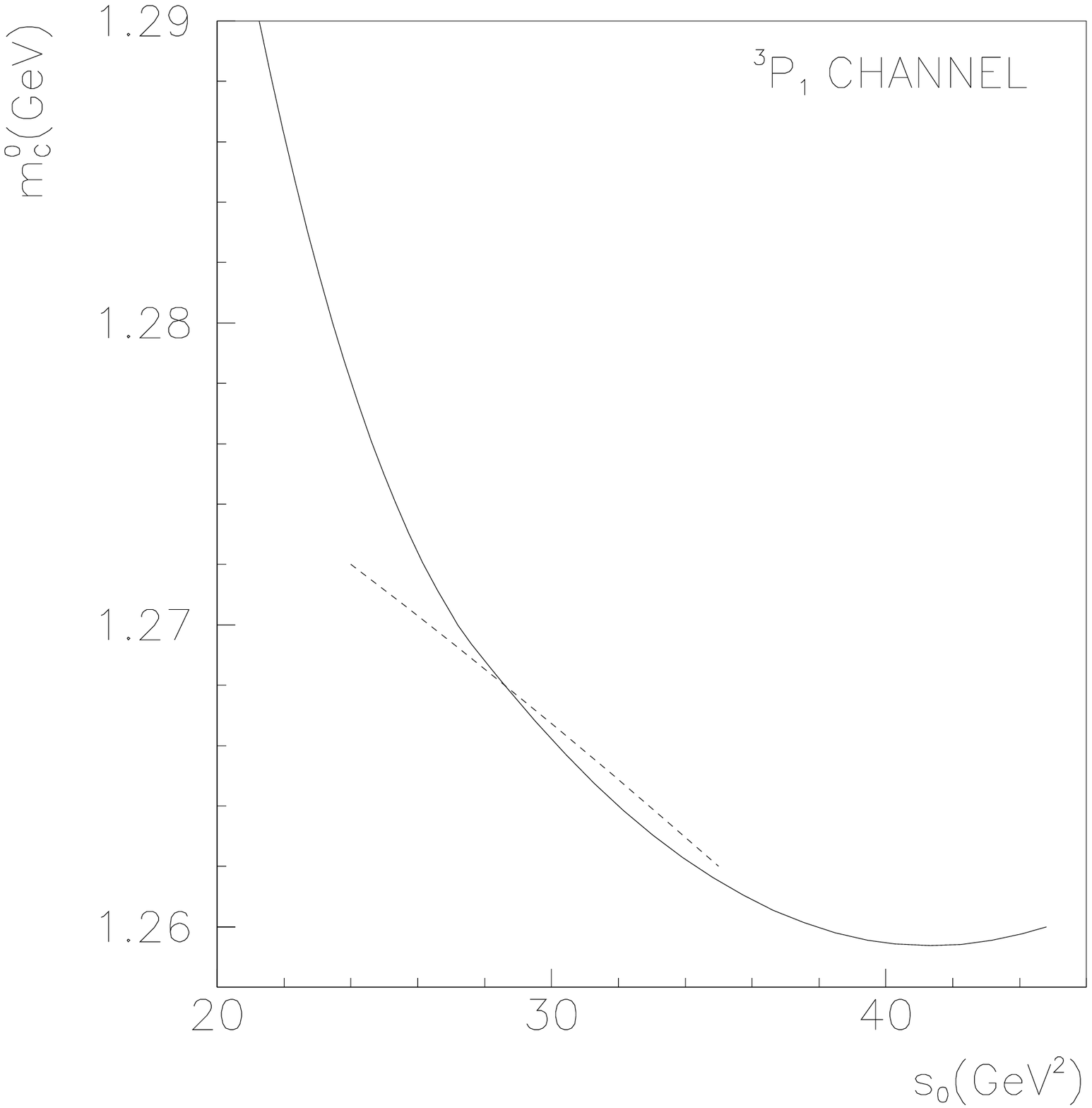,width=7cm} 
\\ 
$b)$ 
}
\caption{$~^3 P_1$ channel: $m_c^0$ vs $s_0$. See caption of fig. 4. Graphs 
$b)$ refer to the last minimum of $D_1$ and present a clear anomaly with 
respect to the others.}
\label{fig:five}
\end{figure}                                              


\begin{thebibliography}{99}

\bibitem {rp} M.Baker, J.S.Ball and F.Zachariasen: Phys. Rev. D {\em 47}, 3021
(1992)

\bibitem {rpp} T.Appelquist et al.: Ann. Rev. Nucl. Part. Sci. 28 ,387 (1978)

\bibitem {li} D.B.Lichtenberg and R.Potting: Phys. Rev. D {\em 46}, 2150 (1992)

\bibitem {ek} A.El-Khadra: Talk given at Tau Charm Factory Workshop, Argonne,
Il, June 21-23 (1995), hep-ph/9509381       

\bibitem {r1}  V.A.Novikov, ~ L.B.Okun, ~ M.A.Shifman, ~ A.I.Vainshtein, ~
M.B.Voloshin and V.I.Zakharov: Phys.Rep. {\em 41}, 1 (1978)

\bibitem {r2} M.A.Shifman, A.I.Vainshtein, M.B. Voloshin and V.I.Zakharov: 
Phys.Lett. B {\em 77}, 80 (1978)

\bibitem {r3} M.A.Shifman, A.I.Vainshtein and V.I.Zakharov: Nucl.Phys. B {\em
147}, 385, 448, 519 (1979)

\bibitem {r4} L.J.Reinders, H.R.Rubinstein and S.Yazaki: Nucl.Phys. B {\em 
186}, 109 (1981) 

\bibitem {r4p} S.Narison: Phys. Lett. B {\em 387}, 162 (1996); Nucl. Phys. B
Proc. Suppl. {\em 54 A}, 238 (1997) 

\bibitem {r4s} C.A.Dominguez and N.Paver: Phys. Lett. B {\em 293}, 197 (1992)

\bibitem {r5} E.DiSalvo and M.Pallavicini: Nucl.Phys. B {\em 427}, 22 
(1994)

\bibitem {r6} E.DiSalvo, M.Pallavicini, E.Robutti and S.Marsano: Phys. Lett. B
{\em 387}, 395 (1996)                   

\bibitem {r6t} J.S.Bell and R.A.Bertlmann: Nucl. Phys. B {\em 177}, 218 (1981)

\bibitem {r6s} E.DiSalvo, M.Pallavicini and E.Robutti: Nucl. Phys. B
Proc. Suppl. {\em 54 A}, 233 (1997) 

\bibitem {r6v} E.DiSalvo: Proc. Int. School of Physics "E.Fermi", Course CXXX,
ed. A.DiGiacomo and D.Diakonov, IOS Press, Amsterdam 1996, p. 469

\bibitem {r7} V.Yu.Borue and S.B.Khokhlachev: Modern Phys. Lett. {\em 3}, 1499
(1988); JETP Lett. {\em 47}, 440 (1988)

\bibitem {rpa}Particle Data Group, R.M.Barnett et al., Review of Particle 
Physics: Phys. Rev. D {\em 54}, 1 (1996)

\bibitem {r8} G.T.Bodwin, E.Braaten and G.P.Lepage: Phys. Rev. D {\em 46},
R1914 (1992); Phys. Rev. D {\em 51}, 1125 (1995)

\bibitem {r9} H.W.Huang and K.T.Chao: Phys. Rev. D {\em 54}, 6850 (1996)

\bibitem {r10} E.Braaten and Yu-Qi Chen: Phys. Rev. D {\em 55}, 7152 (1997)   

\bibitem {r11} R.Barbieri, M.Caffo, R.Gatto and E.Remiddi: Nucl. Phys. B {\em
192}, 61 (1981)

\bibitem {r12} F.Yuan, C.-F.Qiao and K.-T.Chao: Phys. Rev. D {\em 56} 1663
(1997) 

\end{thebibliography}
\end{document}